\documentstyle[preprint,aps]{revtex}
\tighten
\draft
\widetext
\preprint{YITP-00-38}
\bigskip
\bigskip
\begin{document}
\title{``Self-tuning'' and Conformality}
\medskip
\author{Zurab Kakushadze\footnote{E-mail: 
zurab@insti.physics.sunysb.edu}}
\bigskip
\address{C.N. Yang Institute for Theoretical Physics\\ 
State University of New York, Stony Brook, NY 11794}

\date{September 25, 2000}
\bigskip
\medskip
\maketitle

\begin{abstract} 
{}We consider an infinite-volume brane world setup
where a codimension one brane is coupled to bulk 
gravity plus a scalar field with vanishing potential. The latter is protected 
by bulk supersymmetry, which is intact even if brane 
supersymmetry is completely broken as the volume of the extra dimension
is infinite. Within this setup we discuss a flat 
solution with a ``self-tuning'' property, that is, such a solution exists
for a continuous range of values for the brane tension.
This infinite-volume solution is free of any singularities, and has the 
property that the brane cosmological constant is protected by bulk 
supersymmetry. We, however, also point out that consistency of the coupling
between bulk gravity and brane matter 
generically appears to require that the brane world-volume theory be 
conformal.

\end{abstract}

\section{Introduction}

{}In the Brane World scenario the Standard Model gauge and matter fields
are assumed to be localized on  
branes (or an intersection thereof), while gravity lives in a larger
dimensional bulk of space-time 
\cite{early,BK,polchi,witt,lyk,shif,TeV,dienes,3gen,anto,ST,BW}. The volume
of dimensions transverse to the branes is 
automatically finite if these dimensions are compact. On the other hand, 
the volume of the transverse dimensions
can be finite even if the latter are non-compact. In particular, this can be
achieved by using \cite{Gog} warped compactifications \cite{Visser} which
localize gravity on the brane. A concrete
realization of this idea was given in \cite{RS}.

{}Recently it was pointed out in
\cite{DGP0,witten} that, in theories where extra dimensions
transverse to a brane have infinite 
volume\cite{GRS,CEH,DGP0,witten,DGP1,DVALI,DGP,zura,zura1,zura2,DG}, 
the cosmological constant on the
brane might be under control even if brane supersymmetry is completely
broken. The key point here is that even if supersymmetry breaking on the
brane does take place, it will not be transmitted to the bulk as the volume 
of the extra dimensions is infinite \cite{DGP0,witten}. Thus, at least in 
principle, one should be able to control some of the properties of the bulk
with the unbroken bulk supersymmetry. One then can wonder whether bulk 
supersymmetry could also control the brane cosmological constant
\cite{DGP0,witten}.

{}The question whether bulk supersymmetry can control brane cosmological 
constant was addressed in the codimension one brane world scenarios with
an infinite volume extra dimension in \cite{zura,zura1,zura2}.
Thus, in \cite{zura} it was pointed out that if the bulk curvature is
constant, bulk supersymmetry does not control the brane
cosmological constant. However, as was pointed out in \cite{zura1,zura2},
bulk supersymmetry might control the brane cosmological constant
if the bulk curvature is not constant. An example
of such a setup was given in \cite{zura1}. Thus, in the setup of \cite{zura1}
non-constant bulk curvature is parametrized by a bulk scalar $\phi$ which has
a non-trivial profile in the extra dimension due to the fact that the 
corresponding scalar potential depends non-trivially on $\phi$. As was argued
in \cite{zura1}, bulk supersymmetry in this setup indeed controls the brane
cosmological constant. However, as was argued in \cite{zura2}, the brane
world-volume theory in this setup must be conformal in order to have consistent
coupling between bulk gravity and brane matter.

{}In this paper we study a somewhat modified setup. In particular, we consider
a model where the bulk 
scalar potential vanishes. In this case the model has a self-tuning property, 
that is, a flat solution exist for a continuous range of values for the brane
tension. However, unlike the setup of \cite{KSS}, in the model we discuss
in this paper the volume of the extra dimension is infinite, and, in 
particular, the background is free of any naked singularities. Moreover,
since the volume of the extra dimension is infinite, bulk supersymmetry 
protects the self-tuning property. In particular, vanishing of the
bulk scalar potential can be achieved without any fine tuning by imposing 
unbroken $R$-symmetry in the bulk.

{}However, the setup we discuss in this paper also has the property that
consistency of the coupling between brane matter and bulk gravity requires
that the brane world-volume theory be conformal. 
This property is due to the fact that the brane tension in this model is
non-vanishing, and, as was recently pointed out in \cite{zuraRS}, one cannot
gauge away the graviscalar in such backgrounds as the corresponding 
diffeomorphisms are explicitly broken by the presence of such a 
brane. (The brane tension in this model is negative, so the brane must be
placed at a ${\bf Z}_2$ orbifold fixed point to avoid ghosts in the brane 
world-volume theory.) If the brane world-volume theory is not conformal, 
one then expects various terms
involving the graviscalar to be generated in the brane world-volume action
at the quantum level. Such terms, however, 
lead to inconsistencies in the coupling 
between brane matter and bulk gravity.
 
{}The remainder of this paper is organized as follows. In section II
we discuss the setup of this paper. In section III we study 
small fluctuations around the solution in the presence of brane matter
sources, and discuss the requirement that the brane
matter be conformal. Section IV contains concluding remarks.
 
\section{The Model}

{}Consider the
model with the following action (more precisely, here we give the part of
the action relevant for the subsequent discussions):
\begin{equation}\label{action}
 S=-\int_\Sigma d^{D-1} x \sqrt{-{\widehat G}}f +
 M_P^{D-2}
 \int d^D x \sqrt{-G} \left[R-{4\over{D-2}}(\nabla\phi)^2 \right]~.
\end{equation}
For calculational convenience we will keep the number of space-time
dimensions $D$ unspecified.
In (\ref{action}) $M_P$ is the $D$-dimensional (reduced) Planck scale.
The $(D-1)$-dimensional hypersurface $\Sigma$, which we
will refer to as the brane, is the $z=0$ slice of the $D$-dimensional 
space-time,
where $z\equiv x^D$. Next, 
\begin{equation}
 {\widehat G}_{\mu\nu}\equiv{\delta_\mu}^M {\delta_\nu}^N G_{MN}
 \Big|_{z=0}~,
\end{equation} 
where the capital Latin indices $M,N,\dots=1,\dots,D$, while the Greek
indices $\mu,\nu,\dots=1,\dots,(D-1)$. The quantity $f$ is
the brane tension. More precisely, there might be various (massless and/or 
massive) fields (such
as scalars, fermions, gauge vector bosons, {\em etc.}), which we
will collectively denote via $\Phi^i$, 
localized on the brane. Then $f=f(\Phi^i,\nabla_\mu\Phi^i,\dots)$ 
generally depends
on the vacuum expectation values of these fields as well as their derivatives.
In the following we will assume that the expectation values of the $\Phi^i$
fields are dynamically determined, independent of the coordinates 
$x^\mu$, and consistent with $(D-1)$-dimensional general covariance. 
The quantity $f$ is then a constant which we identify 
as the brane tension. 
The bulk fields are given by the metric $G_{MN}$, a single
real scalar field $\phi$, as well as other fields (whose expectation values
we assume to be vanishing) which would appear in a concrete supergravity 
model (for the standard values of $D$). The scalar potential for the field 
$\phi$ is assumed to be vanishing. Note that this can be achieved without
fine-tuning as the bulk is supersymmetric.

{}To proceed further, we will need equations of motion following from the
action (\ref{action}). These are given by:
\begin{eqnarray}
 && {8\over{D-2}}\nabla^2\phi={\sqrt{-{\widehat G}}\over\sqrt{-G}}
 {\widetilde f}_\phi \delta(z)~,\\
 \label{einstein}
 &&R_{MN}-{1\over 2}G_{MN} R
 ={4\over {D-2}}\left[\nabla_M\phi\nabla_N\phi
 -{1\over 2}G_{MN}(\nabla \phi)^2\right]-{1\over 2}
 {\sqrt{-{\widehat G}}\over\sqrt{-G}}{\delta_M}^\mu{\delta_N}^\nu{\widehat
 G}_{\mu\nu}
 {\widetilde f}\delta(z)~.
\end{eqnarray}
Here ${\widetilde f}\equiv f/M_P^{D-2}$, and the subscript $\phi$ indicates
derivative w.r.t. $\phi$ (note that generally $f$ depends on $\phi$, more
precisely, on its value at $z=0$).

{}Here we are interested in studying possible solutions
to these equations which are consistent with $(D-1)$-dimensional general 
covariance. That is, we will be looking for solutions with the warped
metric of the following form:
\begin{equation}\label{warped}
 ds_D^2=\exp(2A)\left[{\widetilde g}_{\mu\nu}dx^\mu dx^\nu +dz^2\right]~,
\end{equation}
where the warp factor $A$ and the scalar field $\phi$, 
which are functions of $z$,
are independent of the coordinates
$x^\mu$, and the $(D-1)$-dimensional metric 
${\widetilde g}_{\mu\nu}$ is independent of $z$. With this 
ans{\"a}tz, we have the following
equations of motion for $\phi$ and $A$:
\begin{eqnarray}\label{phi''d}
 &&{8\over {D-2}}\left[\phi^{\prime\prime}+(D-2)A^\prime\phi^\prime\right]
 =\exp(A){\widetilde f}_\phi\delta(z)~,\\
 \label{phi'A'd}
 &&(D-1)(D-2)(A^\prime)^2-{4\over{D-2}}(\phi^\prime)^2-
 {{D-1}\over{D-3}}{\widetilde \Lambda}=0~,\\
 \label{A''d}
 &&(D-2)\left[A^{\prime\prime}-(A^\prime)^2\right]
 +{4\over {D-2}}(\phi^\prime)^2+{1\over {D-3}}
 {\widetilde \Lambda}=-{1\over 2}\exp(A){\widetilde f}\delta(z)~,
\end{eqnarray}
where a prime indicates derivative w.r.t. $z$.
Next, ${\widetilde \Lambda}$ 
is independent of $x^\mu$ and $z$. In fact, it 
is nothing but the cosmological constant of the $(D-1)$-dimensional manifold,
which is therefore an Einstein manifold, corresponding to the hypersurface
$\Sigma$. Our normalization of ${\widetilde\Lambda}$ is such that
the $(D-1)$-dimensional metric ${\widetilde g}_{\mu\nu}$ satisfies
Einstein's equations:
\begin{equation}
 {\widetilde R}_{\mu\nu}-{1\over 2}{\widetilde g}_{\mu\nu}
 {\widetilde R}=-{1\over 2}
{\widetilde g}_{\mu\nu}{\widetilde\Lambda}~.
\end{equation}
Here we note that in the bulk (that is, for $z\not=0$) one of the
second order equations is automatically satisfied once the first
order equation (\ref{phi'A'd}) as well as the other second order equation are
satisfied. As usual, this is a consequence of Bianchi identities.

\subsection{Bulk Supersymmetry and Brane Cosmological Constant}

{}In the following we will be interested in solutions to the above equations 
of motion such that the volume of the $z$ dimension is infinite. Consistent 
solutions of this type exist for vanishing as well as non-vanishing brane
cosmological constant ${\widetilde\Lambda}$. We will, however, assume that the
bulk is supersymmetric. In fact, as was pointed out in \cite{DGP0,witten},
if the volume of the extra dimension is infinite, bulk supersymmetry is 
intact even if brane supersymmetry is completely broken. Then, as was pointed 
out in \cite{zura1}, if the bulk curvature is not constant\footnote{As was 
pointed out in \cite{zura}, if the bulk curvature is constant, then bulk 
supersymmetry does not control the brane cosmological constant.}, then
bulk supersymmetry might control the brane
cosmological constant. In the above model bulk curvature is not constant as 
long as the $\phi$ field has a non-trivial profile. Then it is not difficult 
to check that bulk supersymmetry indeed controls the brane cosmological 
constant. In fact, this follows from the bulk Killing
spinor equations (following from the requirement that variations of the
superpartner $\lambda$ of $\phi$ and the gravitino $\psi_M$ vanish
under the corresponding supersymmetry transformations), which in such 
backgrounds reduce to:
\begin{eqnarray}\label{killing1}
 &&\phi^\prime=\alpha W_\phi \exp(A)~,\\
 \label{killing2}
 &&A^\prime=\beta W \exp(A)~,
\end{eqnarray}
where $W$ is the superpotential,
\begin{equation}
 \alpha\equiv\eta{\sqrt{D-2}\over 2}~,~~~\beta\equiv-\eta
 {2\over (D-2)^{3/2}}~,
\end{equation}
and $\eta=\pm 1$.

{}Note that the system of equations (\ref{killing1}) and (\ref{killing2})
is compatible with the system of equations (\ref{phi''d}), 
(\ref{phi'A'd}) and (\ref{A''d}) if and only if ${\widetilde\Lambda}=0$,
and 
\begin{equation}\label{supsup}
 W=C\exp(\epsilon\gamma\phi)~,
\end{equation}
where $C$ is a constant, $\epsilon=\pm 1$, and
\begin{equation}
 \gamma\equiv {2\sqrt{D-1}\over{D-2}}~.
\end{equation}
In fact, (\ref{supsup}) is
simply the statement that the 
scalar potential, which is given by the familiar expression 
$V=W_\phi^2-\gamma^2 W^2$, vanishes in the model
defined in (\ref{action}).

{}Thus, bulk supersymmetry (note that a solution to (\ref{killing1}) and
(\ref{killing2}) preserves
$1/2$ of the supersymmetries compared with a flat
solution) is preserved
if and only if the brane cosmological constant vanishes. We therefore
conclude that even if brane supersymmetry is broken, bulk supersymmetry,
which remains unbroken as the volume of the transverse dimension is
infinite, ensures that the brane cosmological constant still vanishes
in the model defined in (\ref{action}).

\subsection{A Self-tuning Solution with Infinite Volume}

{}In this subsection we would like to discuss a solution of (\ref{phi''d}), 
(\ref{phi'A'd}) and (\ref{A''d}) with the self-tuning property. That is, 
this solution has vanishing brane cosmological constant, which, moreover,
exists for a continuous range of values for the brane tension. 

{}Thus, consider the following solution:
\begin{eqnarray}
 &&\phi(z)=\kappa{\sqrt{D-1}\over 2}\ln\left[{{|z|}\over\Delta}+1\right]+
 \phi_0~,\\
 &&A(z)={1 \over{D-2}}\ln\left[{{|z|}\over\Delta}+1\right]+A_0~,
\end{eqnarray}
where $\kappa=\pm 1$, and $\Delta$ is a {\em positive}
quantity, which is related to
the brane tension via
\begin{equation}
 \Delta=-4/{\widetilde f}~.
\end{equation}
That is, in this solution the brane tension $f$ is assumed to be negative.
Moreover, we have an additional condition on $f$ given by ($\epsilon$ and 
$\gamma$ were defined in the previous subsection)
\begin{equation}\label{condi}
 f_\phi(\phi_0)=-\kappa\gamma f(\phi_0)~.
\end{equation}
Note that this solution is non-singular, and the volume of the extra dimension 
is infinite as the integral
\begin{equation}
 \int dz \exp(DA)
\end{equation}
diverges. Moreover, this solution preserves $1/2$ of the original 
supersymmetries in the bulk. The brane cosmological constant in this solution 
is vanishing. Most importantly, such a solution exists for a continuous range
of values for the brane tension (namely, for any negative brane tension) - 
indeed, changing the brane tension amounts to appropriately changing the
parameter $\Delta$. Note that the immaterial integration constant $A_0\equiv
A(0)$ is not determined and can be set to zero, and we will do so in the
following. As to the integration constant $\phi_0\equiv\phi(0)$, it is fixed 
by (\ref{condi}). Thus, the condition (\ref{condi}) is not a fine-tuning, but
rather fixes the value of the scalar field $\phi$ on the brane\footnote{Here
one assumes that $f(\phi)$ is such that there exists $\phi_0$ such that
(\ref{condi}) is satisfied. Otherwise, ${\bf Z}_2$ symmetric solutions with
${\widetilde\Lambda}=0$ do not exist.}.

{}The above solution has a ${\bf Z}_2$ symmetry w.r.t. the reflection 
$z\rightarrow -z$. The reason why we are focusing on this solution is that
we must actually consider the orbifolded version of this solution so that
the geometry of the $z$ dimension is ${\bf R}/{\bf Z}_2$ (and not ${\bf R}$).
The reason for this is that the brane tension is negative, and unless the brane
is an ``end-of-the-world'' brane stuck at the ${\bf Z}_2$ fixed point located 
at $z=0$, the brane world volume theory would suffer from ghosts. In the
following we will therefore adapt the orbifolded version of this solution. 

{}Here the following remark is in order. The self-tuning property of the 
above solution does not imply that there are no solutions with non-zero 
cosmological constant. In fact, it is not difficult to check that, even if
we confine to ${\bf Z}_2$ symmetric solutions (so that we can avoid ghosts
by considering orbifolded versions thereof), solutions
with ${\widetilde\Lambda}\not=0$ do exist. In particular, for such solutions 
the analog of the condition (\ref{condi}) contains the cosmological constant 
${\widetilde \Lambda}$. However, in the above setup what ensures vanishing of
the brane cosmological constant is bulk supersymmetry, which is intact 
even if brane supersymmetry is completely broken as the volume of the extra 
dimension is infinite. The condition (\ref{condi}) then fixes $\phi_0$.

{}Note that in the above discussion it is important that the volume of the 
transverse dimension is infinite. Had it been finite, there would be nothing
there to protect the brane cosmological constant (except if supersymmetry 
is unbroken both on the brane and in the bulk). Thus, consider solutions of the
above type in the cases where the brane tension is positive. Then $\Delta$ 
defined above is negative, and we have a naked singularity at a finite distance
from the brane. Such solutions were originally discussed in \cite{KSS}, where
the space in the $z$ direction was cut off at the singularities, so that the
solution appears to have a finite volume. The self-tuning property, however,
would be lost in finite volume cases. At any rate, as was pointed out in
\cite{COSM}, cutting off the space at singularities arising in such solutions
is not consistent\footnote{Some possibilities for resolving such 
singularities were discussed in \cite{gubser,nilles}.}.

\section{Bulk Gravity and Brane Matter}

{}In this section we would like to study
gravitational interactions between sources localized on the brane. To do
this, let us start from the action (\ref{action}), and study small 
fluctuations of the metric $G_{MN}$ and the scalar field $\phi$, 
which we will denote via $h_{MN}$ and $\varphi$, respectively, around the
self-tuning solution with infinite volume discussed in the previous section. 

{}In the following, instead of metric fluctuations $h_{MN}$, it will be 
convenient to work with ${\widetilde h}_{MN}$ defined via
\begin{equation}
 h_{MN}=\exp(2A) {\widetilde h}_{MN}~.
\end{equation}
It is not difficult to see that in terms of ${\widetilde h}_{MN}$ the
$D$-dimensional diffeomorphisms 
\begin{equation}
 \delta h_{MN}=\nabla_M\xi_N+\nabla_N\xi_M
\end{equation}
are given by the following gauge 
transformations (the capital Latin indices are lowered and raised with the
flat $D$-dimensional Minkowski metric $\eta_{MN}$ and its inverse):
\begin{equation}\label{gauge}
 \delta{\widetilde h}_{MN}=\partial_M {\widetilde\xi}_N+
 \partial_N{\widetilde\xi}_M+2A^\prime\eta_{MN}{\widetilde \xi}_S n^S~. 
\end{equation}
Here for notational convenience we have introduced a unit vector $n^M$
with the following components: $n^\mu=0$, $n^D=1$.

{}Note that, as was pointed out in \cite{zuraRS}, the presence of the 
non-zero tension brane explicitly
breaks the full $D$-dimensional diffeomorphism invariance (\ref{gauge}) to
a smaller subset of gauge transformations given by (\ref{gauge}) with the
restrictions that
\begin{equation}\label{restrict}
 {\widetilde \xi}_D(z=0)=0~,~~~{\widetilde \xi}_\mu^\prime (z=0)=0.
\end{equation}
There is, however, a further restriction on the unbroken gauge transformations
coming from the fact that the brane is stuck at the orbifold fixed point.
This restriction reads:
\begin{equation}
 {\widetilde \xi}^\prime_D (z=0)=0~.
\end{equation}
It is then not difficult to see that if $\rho(z=0)$ is non-zero, then we 
cannot gauge it away even just on the brane, where $\rho\equiv
{\widetilde h}_{DD}$. 
However, we can gauge $A_\mu\equiv {\widetilde h}_{\mu D}$
away everywhere as long as $A_\mu(z=0)=0$, which is indeed the case as we will
see in the next subsection.

\subsection{Equations of Motion} 

{}To proceed further, we need equations of motion for ${\widetilde h}_{MN}$ 
and $\varphi$. Let us
assume that we have matter localized on the brane, and let the corresponding 
conserved energy-momentum tensor be $T_{\mu\nu}$:
\begin{equation}\label{conserved}
 \partial^\mu T_{\mu\nu}=0~.
\end{equation}
The graviton field ${\widetilde h}_{\mu\nu}$ couples to $T_{\mu\nu}$ via
the following term in the action (recall that we have set $A(0)=0$): 
\begin{equation}\label{int}
 S_{\rm {\small int}}=\int_\Sigma d^{D-1} x \left[{1\over 2} T_{\mu\nu}
{\widetilde h}^{\mu\nu}+{8\over{D-2}}\Theta\varphi\right]~,
\end{equation} 
where we have also included the corresponding coupling of $\varphi$ to the
brane matter. Next, starting 
from the action $S+S_{\rm{\small int}}$ we obtain the
following linearized equations of motion for ${\widetilde h_{MN}}$ and
$\varphi$:
\begin{eqnarray}\label{EOMh}
 &&\left\{\partial_S\partial^S {\widetilde h}_{MN} +\partial_M\partial_N
 {\widetilde h}-\partial_M \partial^S {\widetilde h}_{SN}-
 \partial_N \partial^S {\widetilde h}_{SM}-\eta_{MN}
 \left[\partial_S\partial^S {\widetilde h}-\partial^S\partial^R
 {\widetilde h}_{SR}\right]\right\}+\nonumber\\
 &&(D-2)A^\prime\left\{\left[\partial_S {\widetilde h}_{MN} -
 \partial_M {\widetilde h}_{NS}-\partial_N{\widetilde h}_{MS}\right] n^S
 +\eta_{MN}\left[2\partial^R {\widetilde h}_{RS} - \partial_S 
 {\widetilde h}\right] n^S\right\}=\nonumber\\
 &&{8\over {D-2}}\phi^\prime\left[\eta_{MN}\partial_S\varphi n^S-
 \partial_M \varphi n_N-\partial_N\varphi n_M\right]
 -M_P^{2-D} {\widetilde T}_{MN}\delta(z)~,\\ 
 &&\partial_S\partial^S\varphi +(D-2) A^\prime\partial_S\varphi n^S-
 {1\over 2}\phi^\prime 
 \left[2\partial^R {\widetilde h}_{RS} - \partial_S 
 {\widetilde h}\right] n^S=-M_P^{2-D} {\widetilde \Theta}\delta(z)~,
\end{eqnarray}
where ${\widetilde h}\equiv {\widetilde h}_M^M$, ${\widetilde T}_{MN}\equiv
T_{MN}+T^{\rm{\small brane}}_{MN}$, ${\widetilde \Theta}\equiv\Theta+
\Theta^{\rm{\small brane}}$. Here $T^{\rm{\small brane}}_{MN}$ and 
$\Theta^{\rm{\small brane}}$ are the corresponding brane
contributions (which are linear in ${\widetilde h}_{MN}$ and $\varphi$)
coming from the first term in (\ref{action})\footnote{If the brane 
world-volume theory is not
conformal, then we can {\em a priori} expect additional contributions 
arising due to quantum corrections. We will discuss effects of
such contributions in the next subsection, which will lead us to the
conclusion that to maintain consistent
coupling between bulk gravity and brane matter the brane world-volume theory
should be conformal.}. Thus, we have:
\begin{eqnarray}
 &&T^{\rm{\small brane}}_{\mu\nu}=-\eta_{\mu\nu}\left[f_\phi \varphi+
 {1\over 2} f\rho\right]~,\\
 &&T^{\rm{\small brane}}_{\mu D}=A_\mu f~,\\
 &&T^{\rm{\small brane}}_{DD}=0~,\\  
 &&\Theta^{\rm{\small brane}}=-{{D-2}\over 8}\left[f_{\phi\phi}\varphi+
 {1\over 2}f_\phi\rho\right]~.
\end{eqnarray}
Note that {\em a priori} $\partial^\mu {\widetilde T}_{\mu\nu}\not=0$.

{}It is not difficult to see that the r.h.s. of the $(\mu D)$ component
of (\ref{EOMh}) does not contain any terms with the second derivative w.r.t.
$z$. This then implies that, to have a consistent solution, we must have
$A_\mu(z=0)=0$. As we have already mentioned, we can then
gauge $A_\mu$ away
everywhere, and we will adapt this gauge in the following. 
We then have the following equations of motion (the Greek indices are
lowered and raised with the flat $(D-1)$-dimensional Minkowski metric
and its inverse):
\begin{eqnarray}\label{EOM1}
 &&\left\{\partial_\sigma\partial^\sigma 
 H_{\mu\nu} +\partial_\mu\partial_\nu
 H-\partial_\mu \partial^\sigma H_{\sigma\nu}-
 \partial_\nu \partial^\sigma H_{\sigma\mu}-\eta_{\mu\nu}
 \left[\partial_\sigma\partial^\sigma H-\partial^\sigma\partial^\rho
 H_{\sigma\rho}\right]\right\}+\nonumber\\
 &&\left\{H_{\mu\nu}^{\prime\prime}-\eta_{\mu\nu}H^{\prime\prime}+
 (D-2)A^\prime\left[H_{\mu\nu}^\prime-\eta_{\mu\nu}H^\prime\right]\right\}+
 \nonumber\\
 &&\left\{\partial_\mu\partial_\nu\rho-\eta_{\mu\nu}
 \partial_\sigma\partial^\sigma 
 \rho+\eta_{\mu\nu}(D-2)A^\prime\rho^\prime
 \right\}-{8\over{D-2}}\eta_{\mu\nu}\phi^\prime
 \varphi^\prime=-M_P^{2-D} {\widetilde T}_{\mu\nu} \delta(z)~,\\ 
 \label{EOM2} 
 &&\left[\partial^\mu H_{\mu\nu}-\partial_\nu H\right]^\prime +(D-2)A^\prime 
 \partial_\nu\rho-{8\over{D-2}}\phi^\prime\partial_\nu\varphi=0~,\\
 \label{EOM3}
 &&-\left[\partial^\mu\partial^\nu H_{\mu\nu}-\partial^\mu\partial_\mu H\right]
 +(D-2) A^\prime H^\prime -{8\over{D-2}}\phi^\prime\varphi^\prime=0~,\\
 \label{EOM4}
 &&\partial^\mu\partial_\mu\varphi+\varphi^{\prime\prime}+(D-2)A^\prime
 \varphi^\prime
 +{1\over 2} \phi^\prime \left[H^\prime-\rho^\prime\right]=
 -M_P^{2-D}{\widetilde\Theta}\delta(z)~,
\end{eqnarray}
where $H\equiv H_\mu^\mu$. Here we note that, as usual, not all of the above
equations are independent.

{}The above system of equations can be analyzed much along the lines of 
\cite{zura2,zuraRS}, so we will skip the details and simply give the answer.
Thus, after some straightforward (albeit tedious) computations one can 
show that for a consistent solution to the above equations to exist we must
have\footnote{Here we note that this condition is analogous to that 
arising in the setup studied in \cite{zura2}.} ($\kappa=\pm 1$ was defined
in section II)
\begin{equation}
 {\widetilde\Theta}=-{\kappa\over {4\sqrt{D-1}}}{\widetilde T}~,
\end{equation}
where $T\equiv T_\mu^\mu$.
The corresponding solution is then given by:
\begin{eqnarray}
 &&\rho=H~,~~~\varphi=\kappa{{D-2}\over{4\sqrt{D-1}}} H~,\\
 \label{prop}
 &&H_{\mu\nu}(p,z)=M_P^{2-D}\left[T_{\mu\nu}(p)-{1\over{D-2}} 
 \eta_{\mu\nu} T(p)
 \right]\Omega(p,z)~,
\end{eqnarray}
where
$\Omega$ is the solution to the following equation 
($p^2\equiv p^\mu p_\mu$)
\begin{equation}
 \Omega^{\prime\prime}(p,z)+(D-2)A^\prime\Omega^\prime(p,z)-p^2\Omega(p,z)=-
 \delta(z)
\end{equation}
subject to the boundary conditions (for $p^2>0$)
\begin{equation}
 \Omega(p,z\rightarrow\pm \infty)=0~.
\end{equation}
In the above expressions we have performed a Fourier transform for the 
coordinates $x^\mu$ (the corresponding momenta are $p^\mu$), and Wick
rotated to the Euclidean space (where the propagator is unique).

{}Note that for the above solution we have $T^{\rm{\small brane}}_{\mu\nu}=0$,
so that ${\widetilde T}_{\mu\nu}=T_{\mu\nu}$. The tensor structure of the
graviton propagator following from (\ref{prop}) is that of the $D$-dimensional
massless graviton (and not of the $(D-1)$-dimensional one), which is in accord
with the fact that the 
volume of the extra dimension is infinite. As to the scalar
degrees of freedom $\rho$ and $\varphi$, they couple to the trace of the
energy-momentum tensor on the brane, that is, the corresponding couplings are
non-vanishing as long as the brane world-volume theory is not conformal.

\subsection{Conformality}

{}The fact that $\rho$ cannot be gauged away has important implications.
Thus, suppose that the brane world-volume theory is not conformal. Then
quantum corrections on the brane will 
generically generate various terms in the brane world-volume \cite{DGP,DG}. 
These terms 
are {\em a priori} arbitrary except that they must respect the gauge symmetries
of the system. On the brane these symmetries are given by the 
$(D-1)$-dimensional diffeomorphism under which $\rho$ and $\varphi$ do not
transform. Thus, for instance, 
consider such terms that do not contain any derivatives. The most general 
corrections of this type can be written as follows:
\begin{equation}
 S_1 =-M_*^{D-1} \int_\Sigma dx^{D-1} F(\rho,\varphi)~,
\end{equation}   
where $F(\rho,\varphi)$ is some dimensionless function of $\rho$ and $\varphi$,
and $M_*$ denotes a mass scale in the brane world-volume theory which 
determines the size of quantum corrections 
(this can be the cut-off scale of the
brane world-volume theory, or the mass scale of some heavy fields localized 
on the brane). Note that the $\rho$-independent part of $F(\rho,\varphi)$, that
is, $F(0,\varphi)$ corresponds to renormalizing $f(\phi)$ in the first term
in (\ref{action}). Due to the self-tuning property of the solution we are 
discussing here, such a renormalization does not affect the qualitative 
features
of the solution as long as the equation
\begin{equation}
 {\widehat f}_\phi(\phi_0)=-\kappa\gamma {\widehat f}(\phi_0)~,
\end{equation}
where ${\widehat f}$ is the renormalized counterpart of $f$,
has a solution such that
${\widehat f}(\phi_0)$ is negative. However, the $\rho$-dependent part of
$F(\rho,\varphi)$, which is generically expected to be non-vanishing as long as
the brane world-volume theory is not conformal, introduces an inconsistency 
into the system. Thus, note that the r.h.s. of (\ref{EOM3}) now contains a
non-vanishing source term proportional to $\delta(z)$. Since the l.h.s. of 
this equation does not contain any second derivatives w.r.t. $z$, we conclude
that we no longer have a consistent solution.

{}Thus, as we see, the fact that $\rho$ cannot be gauged away on the brane
necessitates
fine-tuning at the quantum level to preserve consistent coupling between
bulk gravity and brane matter as long as the brane world-volume theory is not 
conformal. On the other hand, if we assume that the brane world-volume theory 
is conformal, the undesirable terms discussed above will not be generated. 
Indeed, in this case we have $T=0$, and  
\begin{eqnarray}
 &&\rho=\varphi=0~,\\
 &&H_{\mu\nu}(p,z)=M_P^{2-D}T_{\mu\nu}(p)\Omega(p,z)~.
\end{eqnarray}
This then implies that the coupling of the scalar field $\varphi$ to the
brane matter also vanishes: $\Theta=0$. Quantum corrections then will not 
generate any dangerous terms (including those containing derivatives), and the
background as well as the coupling between brane matter and bulk gravity 
should remain consistent at the quantum level.

\section{Remarks}

{}The requirement that the brane world-volume theory be conformal has already
appeared in a somewhat different context in \cite{zura1,zura2}. There too
conformality is required to maintain consistent coupling between 
brane matter and bulk gravity. In \cite{zura1,zura2} as well as in the model
discussed in this paper it is important that the extra dimension has infinite 
volume (and also that the bulk is supersymmetric). Thus, as was recently 
pointed out in \cite{zuraRS}, similar inconsistencies are 
expected\footnote{Some of the terms in the brane world-volume action 
that would lead to such inconsistencies in this context where discussed in
\cite{zuraRS}, and such terms indeed seem to be present \cite{iglesias}.} 
to arise at
the quantum level in, say, the original Randall-Sundrum model \cite{RS}.
There, however, the brane world-volume theory cannot be conformal as long as 
gravity is localized\footnote{As was pointed out in \cite{COSM,olindo}, gravity
in such backgrounds is generically expected to be delocalized due to higher
curvature bulk contributions.}.

{}It would be interesting to understand whether consistency of the coupling
between brane matter and bulk gravity imposes any constraints on the
brane world-volume theory in the cases of higher codimension brane world
scenarios with infinite volume extra dimensions \cite{DG}. In particular, if
in some cases the brane world-volume theory must be conformal, it would be
interesting to understand if there is a relation to \cite{BKV}.

\acknowledgments

{}I would like to thank Gia Dvali and Gregory Gabadadze for discussions.
This work was supported in part by the National Science Foundation.
I would also like to thank Albert and Ribena Yu for financial support.

\end{document}